\documentclass[aps,prl,twocolumn,superscriptaddress]{revtex4-2}  
\usepackage{amsfonts}
\usepackage{bbding}
\usepackage{amssymb}
\usepackage[title]{appendix}
\usepackage{color}
\usepackage{amsmath}
\usepackage{graphicx}
\usepackage{subfigure}
\usepackage{txfonts}
\usepackage{bm}
\usepackage[normalem]{ulem}
\usepackage{natbib}
\usepackage{calrsfs}
\usepackage[colorlinks,linkcolor=blue,citecolor=blue,urlcolor=blue]{hyperref}
\usepackage{orcidlink}
\preprint{APS/123-QED}
\usepackage{tikz}
\usetikzlibrary{graphs, positioning, quotes} 
\usetikzlibrary{matrix}
\newcommand{\resetappendix}{%
    \appendix
    \renewcommand{\thesection}{A\arabic{section}}
    \renewcommand{\theequation}{A\arabic{equation}}
    \renewcommand{\thefigure}{A\arabic{figure}}
    \renewcommand{\thetable}{A\arabic{table}}
    \setcounter{section}{0}
    \setcounter{equation}{0}
    \setcounter{figure}{0}
    \setcounter{table}{0}
}

\DeclareMathAlphabet{\pazocal}{OMS}{zplm}{m}{n}
\newcommand{\Ls}{\pazocal{L}}

\newcommand{\Ds}{\pazocal{D}}
\newcommand{\Zs}{\pazocal{Z}}

\begin{document}

\title{Canonical Quantum Mpemba Effect in a Dissipative Qubit}

\author{Xingli Li\orcidlink{0000-0003-2339-6557}}
\affiliation{Department of Physics, The Chinese University of Hong Kong, Shatin, New Territories, Hong Kong, China}

\author{Yan Li\orcidlink{0009-0007-7218-133X}}
\affiliation{Department of Physics, The Chinese University of Hong Kong, Shatin, New Territories, Hong Kong, China}

\author{Yangqian Yan\orcidlink{0000-0002-3237-5945}}
\email{yqyan@cuhk.edu.hk}
\affiliation{Department of Physics, The Chinese University of Hong Kong, Shatin, New Territories, Hong Kong, China}
\affiliation{State Key Laboratory of Quantum Information Technologies and Materials, The Chinese University of Hong Kong, Hong Kong SAR, China}
\affiliation{The Chinese University of Hong Kong Shenzhen Research Institute, 518057 Shenzhen, China
}

\begin{abstract}
The Mpemba effect, where a hotter system cools faster than a colder one under otherwise identical conditions, has been extensively studied in classical systems.
In this work, we present the quantum analogue of the Mpemba effect using a dissipative qubit, which is referred to as the {\it canonical quantum Mpemba effect}.
We demonstrate that, under the identical conditions, the relaxation dynamics of a qubit initialized in a thermal state with a higher temperature can be exponentially faster than that of a colder thermal state. 
Strikingly, this acceleration is determined solely by the initial temperature of the system, independent of other parameters. 
The relaxation is confirmed to be a genuine cooling process via the effective steady-state temperature, mirroring its classical counterpart. 
Last, we propose a practical classical–quantum hybrid algorithmic quantum circuit to realize this effect using superconducting qubits experimentally. 
\end{abstract}
\date{\today}

\maketitle

{\it Introduction.}--- 
Anomalous relaxation dynamics in non-equilibrium systems have attracted significant attention, with the Mpemba effect~\cite{Mpemba1969} being a well-known example. 
This phenomenon, where a system prepared at a higher initial temperature can relax faster than one at a lower temperature, has been extensively analyzed in classical systems, revealing its dependence on initial states and system properties.
Recently, this counterintuitive phenomenon has been extended to the quantum realm~\cite{Ares2025,Lapolla2020,Yamashika2025,Klobas2025,Alyuruk2025,Summer2025,VanVu2025a,Teza2025}, where studies have investigated anomalous relaxation dynamics in both closed~\cite{Turkeshi2025,Ares2023,Liu2024,Rylands2024,Yamashika2024,Yu2025c,Yamashika2025a,Bhore2025,Ares2025a,Chang2024,Gibbins2025,Liu2024p,Sugimoto2025,Yu2025PRB} and open quantum systems~\cite{Carollo2021,Nava2024,Chatterjee2023,Strachan2025,Medina2025,Tejero2025,Wang2024m,Zhou2023b,Chatterjee2024,Moroder2024,Longhi2024a,Longhi2024b,Dong2025c,Caceffo2024,Boubakour2024,Liu2024g,Su2025,Zhao2025a,Longhi2025a,Ma2025b,Nava2025,Qian2024a,Graf2024,Ramon-Escandell2025,Wang2024n,Bao2025,Wei2025c}.
Furthermore, experimental demonstrations on quantum simulation platforms~\cite{Zhang2025,AharonyShapira2024,Joshi2024a,Xu2025f} have confirmed its feasibility, highlighting the universality of the Mpemba effect in quantum systems.

The manifestation of the Mpemba effect differs between closed and open quantum systems. 
In closed quantum systems, it refers to the anomalous phenomenon where a more asymmetric initial state restores the system's symmetry faster than a less asymmetric one~\cite{Turkeshi2025,Ares2023,Liu2024,Yamashika2024,Rylands2024,Yu2025c,Yamashika2025a}.
By contrast, in open quantum systems, the Mpemba effect is defined as an exponentially accelerated relaxation to the steady state~\cite{Carollo2021,Moroder2024,Medina2025,Chatterjee2023,Nava2024,Strachan2025,Tejero2025,Wang2024m,Zhou2023b,Chatterjee2024}. 
This acceleration is achieved through the careful engineering of initial states via unitary transformations, and this state engineering exploits a key mechanism in open quantum systems: suppression of contributions from the Liouvillian's slowest-decaying eigenmode~\cite{Carollo2021}.

\begin{figure}[!htpb]
    \centering
    \includegraphics[width=0.48\textwidth]{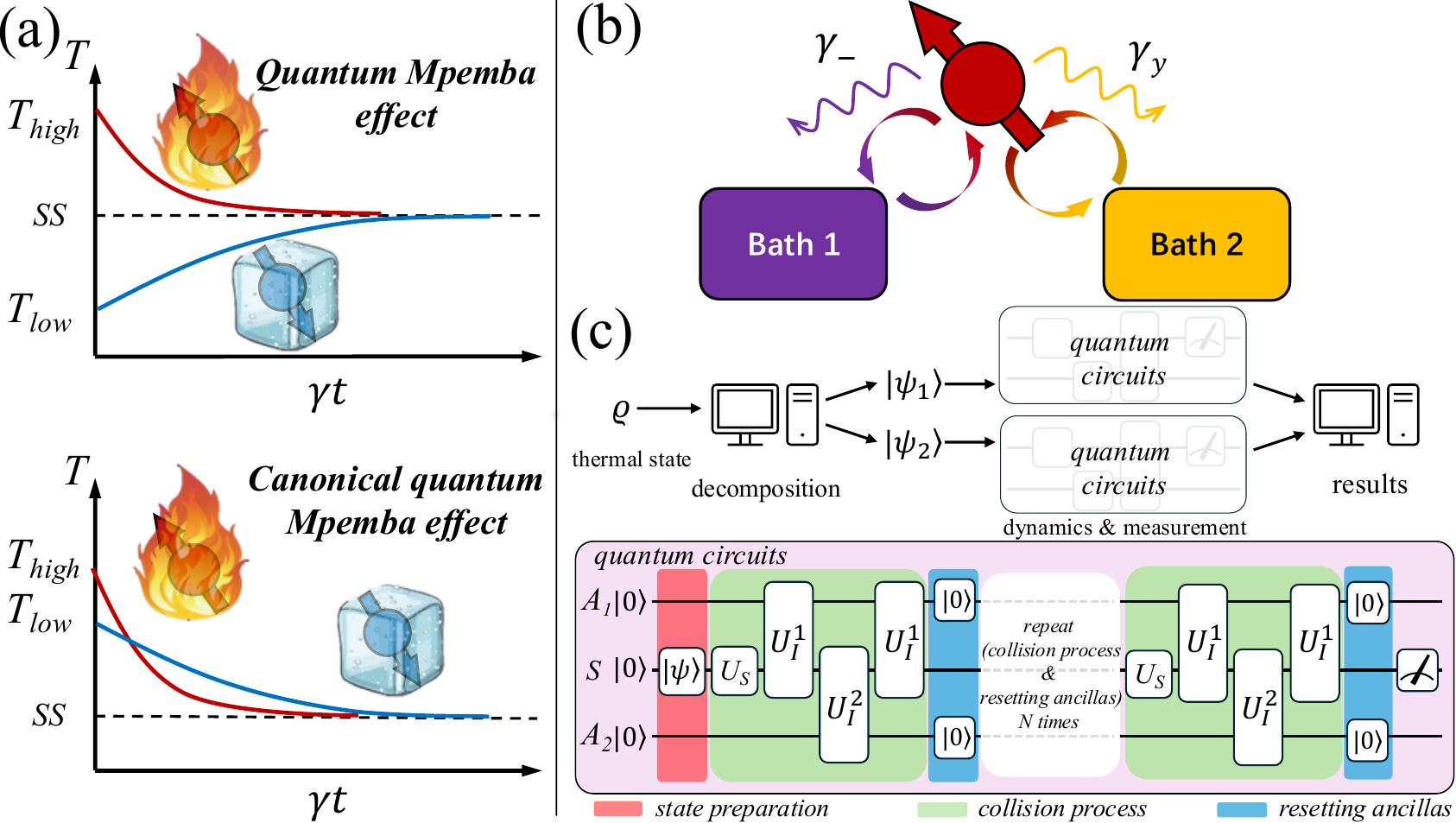}    
    \caption{\label{Fig:Schematic} (a) Schematic of the canonical quantum Mpemba effect. Top panel: traditional quantum Mpemba effect focus on how fast the initial state relax to the steady state. Bottom panel: the canonical quantum Mpemba effect considered in this paper in addition requires both process to be cooling processes.
    (b) Schematic of the model: a qubit coupled to two independent baths, each inducing a distinct dissipation channel with decay rates $\gamma_{-}$ and $\gamma_{y}$. 
    (c) Schematic of the canonical quantum Mpemba effect implemented on a quantum circuit, demonstrating our proposed classical-quantum hybrid algorithm with a thermal initial state.}
\end{figure}

To set the stage, the top panel of Fig.~\ref{Fig:Schematic}(a) depicts a case where the hotter thermal state reaches the steady state faster than its colder counterpart, seemingly fulfilling the usual working definition of a quantum Mpemba effect: a hotter state relaxes faster than a colder one under otherwise identical conditions. We refer to this minimal criterion as the traditional quantum Mpemba effect. However, this scenario mixes cooling and heating trajectories—the hotter state cools while the colder state heats toward the steady state—and thus differs in thermodynamic directionality from the classical Mpemba effect, which involves only cooling processes. In addition, in open quantum systems the steady-state temperature $T_{\text{SS}}$ need not coincide with any bath temperature, especially in the presence of multiple baths, coherence, or other non-equilibrium features; an effective steady-state temperature must therefore be defined. These observations motivate a stronger, thermodynamically consistent formulation that we call the {\it canonical quantum Mpemba effect}, characterized by two requirements: (i) {\it Thermodynamic directionality}—both relaxation trajectories are cooling, with $T_{\text{high}}>T_{\text{low}}>T_{\text{SS}}$; and (ii) {\it Exponential acceleration}—the hotter initial thermal state exhibits exponential speedup via suppression of the slowest Liouvillian mode. The lower panel of Fig.~\ref{Fig:Schematic}(a) illustrates this canonical regime.

In this Letter, we use a simple yet fundamental open quantum system to demonstrate the canonical quantum Mpemba effect: a qubit coupled to two independent thermal baths, and each bath engineering a distinct dissipation channel [Fig.~\ref{Fig:Schematic}(b)].
Our key findings are as follows:
(1) By defining the steady-state effective temperature $T_{\text{SS}}$ as the temperature of thermal state that minimizes the trace distance to the steady state, we identify relaxation dynamics that fulfill the criteria for the canonical quantum Mpemba effect, wherein a higher initial temperature suppresses the slowest relaxation mode, thus enabling an exponentially accelerated cooling process.
(2) We propose an experimental scheme to realize the canonical quantum Mpemba effect based on a quantum collision model, which can be implemented with current quantum simulation technologies. The dissipative dynamic from a thermal initial state is emulated by a classical-quantum hybrid algorithm [Fig.~\ref{Fig:Schematic}(c)], and the effectiveness of this approach is validated on a superconducting quantum circuit.

{\it Model.}---
We consider a qubit system with the Hamiltonian is $H_{S}=\omega_{y}\sigma^{y}_{S}+\omega_{z}\sigma^{z}_{S}$, where $\sigma^{y(z)}_{S}$ are the Pauli operators, $\omega_{y(z)}$ are the frequencies ($\hbar=k_{B}=1$ hereinafter).
The initial states are thermal states: $\varrho_{\rm th}[T]=e^{-\beta H}/\Zs$, where $\Zs=\text{Tr}[e^{-\beta H}]$ is the partition function and $\beta=1/k_{B}T$ is the inverse temperature.
The dynamics of this system are governed by the Lindblad master equation~\cite{Breuer2010} 
\begin{equation}
 \frac{d}{dt}\varrho=-i[H_{S},\varrho]+\gamma_{-}\Ds_{\varrho}[\sigma_{-}]+\gamma_{y}\Ds_{\varrho}[\sigma_{y}],
\label{Eq:ME}   
\end{equation}
where $\Ds_{\varrho}[o]=o\varrho o^{\dagger}-\frac{1}{2}(o^{\dagger}o\varrho+\varrho o^{\dagger}o)$ denotes the dissipator, and $\sigma_{-}$ and $\sigma_{y}$ are the jump operators with respective rates $\gamma_{-}$ and $\gamma_{y}$. 
We set $\gamma_{-}=\gamma_{y}=\gamma$ for simplicity.
The Lindblad master equation describes open quantum system dynamics via the Liouvillian $\Ls$~\cite{Carollo2021}, defined by the linear map $d\varrho/dt=\Ls[\varrho]$. 
The Liouvillian can be represented in matrix form and is diagonalizable, with the right eigenoperators are $\Ls r_{k}=\lambda_{k}r_{k}$ and left eigenoperators are $\Ls^{\dagger}l_{k}=\lambda_{k}^{*}l_{k}$, where the eigenvalues $\lambda_{k}$ are complex numbers with $\text{Re}(\lambda_{k})\leq0,\forall k$~\cite{Breuer2010}.
We order $\lambda_{k}$ by descending real part: $\text{Re}(\lambda_{d^2})\leq\dots\leq\text{Re}(\lambda_{2})\leq\text{Re}(\lambda_{1})=0$, with $d$ being the dimension of the Hilbert space.
\begin{figure}[!htpb]
    \centering
    \includegraphics[width=0.485\textwidth]{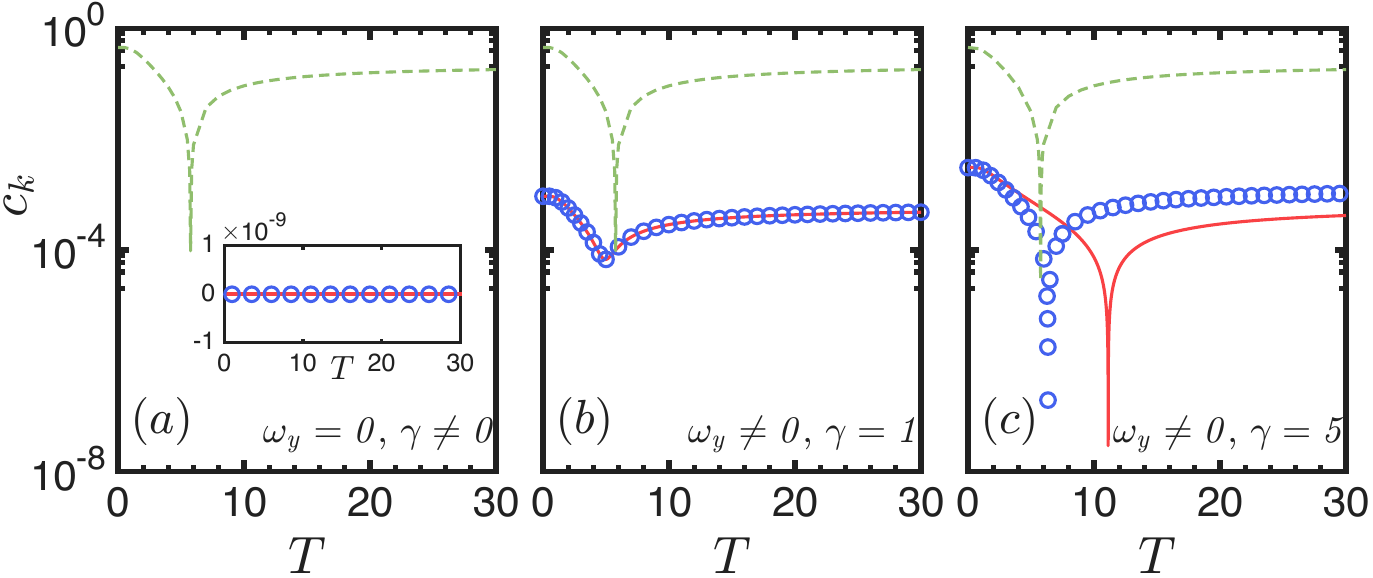}    
    \caption{\label{Fig:ck} 
    The overlaps $c_k$ between the eigenmodes $l_k$ and the initial thermal state as a function of its temperature $T$. 
    The overlaps $c_2$ (red solid), $c_3$ (blue circles), and $c_4$ (green dashed) are plotted, with $\omega_z=2$ fixed throughout. (a) In the absence of off-diagonal term $\omega_{y}$, the inset shows that $c_2$ and $c_3$ are always zero.
    (b) For finite off-diagonal term $\omega_{y}=0.01$ and weak dissipation $\gamma=1$, $c_2$ and $c_3$ overlap and are smooth. (c) For finite off-diagonal term $\omega_{y}=0.01$ and strong dissipation $\gamma=5$,
    the sharp dip of $c_{2}$ at T=11.13 indicates the potential for the canonical quantum Mpemba effect.
    }
\end{figure}
Therefore, the general evolution of a given initial state $\varrho_{\text{ini}}$ takes the form
\begin{equation}
\varrho(t)=e^{t\Ls}[\varrho_{\text{ini}}]=\varrho_{\text{SS}}+\sum_{k=2}^{d^{2}}\text{Tr}[l_{k}\varrho_{\text{ini}}]r_{k}e^{\lambda_{k}t},  
\label{Eq:SpecRho} 
\end{equation}
where $\varrho_{\text{SS}}=r_{1}/\text{Tr}[r_{1}]$ denotes the steady state, $c_{k}=|\text{Tr}[l_{k}\varrho_{\text{ini}}]|$ is the overlap between the initial state $\varrho_{\text{ini}}$ and left eigenoperators $l_{k}$. 
In long-time limit ($t\to\infty$), the relaxation is dominated by the slowest eigenmode, thus the timescale for relaxation is $\tau=|\text{Re}(\lambda_{2})|^{-1}$.
Equation~\eqref{Eq:SpecRho} implies that suppressing the slowest mode accelerates the relaxation process, which requires its overlap $c_{2}$ to approach zero. 
This means the temperature dependence of the overlaps $c_{k}$ indicates whether relaxation is accelerated.
To this end, we discuss the behaviors of $c_{k}$ versus the initial thermal state's temperature $T$.

Figure~\ref{Fig:ck} shows the variation of $c_{k}$ as a function of $T$. 
This encompasses the three scenarios of Eq.~(1), defined by the off-diagonal term in the Hamiltonian $\omega_y$ (either zero or finite) and the dissipation strength $\gamma$ (either weak or strong).
We find that in the absence of the off-diagonal term ($\omega_y=0$) [Fig.~\ref{Fig:ck}(a)], the behaviors of $c_k$ are independent of the dissipation strength.
Moreover, $c_2$ and $c_3$ remain zero at all times, which implies the dynamic is solely governed by the fastest-decaying mode, and the relaxation cannot be accelerated. 
This behavior can be explained by the diagonal Hamiltonian, which prepares an initial thermal state devoid of the coherence. 
The jump operator $\sigma_{-}$ then drives relaxation at the fastest possible rate, leaving no room for acceleration.
Introducing a slight off-diagonal term ($\omega_y=0.01$) dramatically alters the scenarios [Figs.~\ref{Fig:ck}(b)-(c)]. 
Under weak dissipation [Fig.~\ref{Fig:ck}(b)], although $c_2$ and $c_3$ vary with temperature, $c_2$ never approaches zero, precluding acceleration. 
In contrast, under strong dissipation, $c_2$ plummets near $T \approx 11.13$, suggesting a potential canonical Mpemba effect. 
Confirmation, however, requires verifying that the steady-state effective temperature lies below this potential acceleration temperature.

{\it The steady-state effective temperature.}--- Since the steady state of an open system cannot always be characterized by the Gibbs ensemble, $\varrho_{\text{SS}}\not\propto e^{-\beta H_{S}}$~\cite{Trushechkin2022,Sun2024}, we determine $T_{\text{SS}}$ by the thermal state with the shortest distance to the steady state,
\begin{equation}
    T_{\text{SS}}=\arg\min_{T}\left\{D(\varrho_{\text{SS}},\varrho_{\text{th}}[T])\right\},
\end{equation}
where $D(\varrho_{a},\varrho_{b})$ is trace distance~\cite{Nielsen2010} defined as
\begin{equation}
    D(\varrho_{a},\varrho_{b})=\frac{1}{2}\text{Tr}[\sqrt{(\varrho_{a}-\varrho_{b})^{\dagger}(\varrho_{a}-\varrho_{b})}].
\end{equation}
Trace distance is a metric that quantifies the distinguishability between two states, ranging from 0 (indistinguishable) to 1 (perfectly distinguishable). 
Furthermore, we consider quantum relative entropy~\cite{Nielsen2010} as an alternative measure to quantify the difference between two states, which is defined as
\begin{equation}
    S(\varrho_{a}||\varrho_{b})=\text{Tr}[\varrho_{a}\log\varrho_{a}-\varrho_{a}\log\varrho_{b}].
\end{equation} 
Quantum relative entropy reflects the informational difference between two states. 
It is asymmetric and quantifies how much one state deviates from another. 

\begin{figure}[!tpb]
    \centering
    \includegraphics[width=0.475\textwidth]{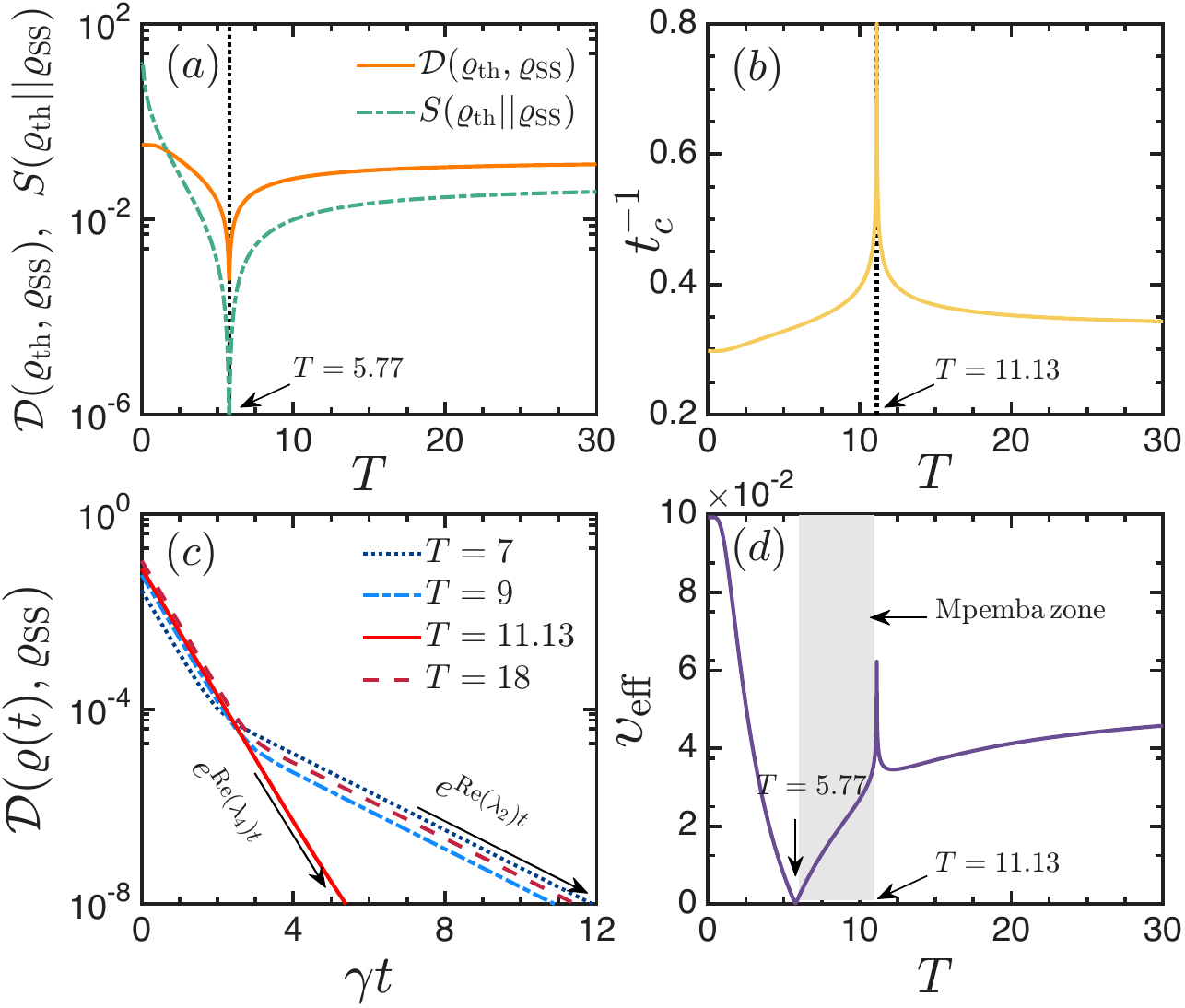}    
    \caption{\label{Fig:Ttim} 
    (a) Trace distance (orange solid line) and quantum relative entropy (green solid line) between the steady state and the thermal states as a function of temperature $T$. 
    The dashed line indicates that when $T\approx 5.77$, the distance between $\varrho_{\text{SS}}$ and $\varrho_{\text{th}}[T]$ is minimized; i.e., the steady state is best described by a thermal state at $T\approx5.77$ 
    (b) The inverse critical time $t_{c}^{-1}$ to reach the steady state as a function of the initial-state temperature $T$. 
    The dashed line indicates that initial thermal state at $T\approx11.13$ reaches the steady state the fastest. 
    (c) Trace distance between the time-evolved state $\varrho(t)$ and the steady state $\varrho_{\text{SS}}$ as a function of the dimensionless time $\gamma t$ for different initial temperatures.
    (d) The effective velocity $\upsilon_{\rm eff}$, i.e., the ratio of initial trace distance to
the critical time,  as a function of the initial-state temperature $T$. 
    The gray zone with $T\in[5.77,11.13]$ indicates the region where the canonical quantum Mpemba effect occurs. 
    }
\end{figure}

Figure~\ref{Fig:Ttim}(a) shows the trace distance (orange solid line) and quantum relative entropy (green solid line) between the steady state $\varrho_{\text{SS}}$ and thermal states $\varrho_{\text{th}}[T]$ across a range of temperatures $T$.
We find that both measures show a deep valley, indicating that the thermal state at this temperature is closest to the steady state.
The effective temperature of the steady state is thus determined as $T_{\text{SS}}\approx 5.77$.


{\it Quantum Mpemba effect.}--- After determining the steady-state effective temperature, we now check whether the canonical quantum Mpemba effect occurs.
In Fig.~\ref{Fig:Ttim}(b), we show the inverse critical time $t_{c}^{-1}$ as a function of temperature.
The critical time $t_{c}$ is defined as the time when trace distance between the time-evolved state $\varrho(t)$ and $\varrho_{\text{SS}}$ reaches a threshold value, which we set as $D(\varrho(t),\varrho_{\text{SS}})\sim10^{-8}$.
The yellow solid line in Fig.~\ref{Fig:Ttim}(b) shows that the inverse critical time $t_{c}^{-1}$ begins at approximately $0.298$ at the zero temperature and increases with rising temperature. 
When the temperature approaches $T\approx 11.13$, the inverse critical time reaches a maximum value.
This indicates that the system is accelerated to reach the steady state at this temperature, as predicted by the overlap $c_{2}$ in Fig.~\ref{Fig:ck}(d).

To better demonstrate the relaxation acceleration, we show the trace distance between the time-evolved state $\varrho(t)$ and the steady state $\varrho_{\text{SS}}$ as a function of the dimensionless time $\gamma t$ for different initial temperatures $T$ in Fig.~\ref{Fig:Ttim}(c).
We present four specific temperatures: $T=7$ (dark blue dotted line), $T=9$ (blue dashdotted line), $T=11.13$ (red solid line), and $T=18$ (dark red dashed line).
During the initial dynamics, the trace distance decreases exponentially as $D(\varrho(t),\varrho_{\text{SS}})\sim e^{\text{Re}(\lambda_{4})t}$, independent of the initial distance between the thermal states (at different temperatures) and the steady state.
Then, when the time approaches $\gamma t \approx 2.5$, the behaviors diverge.
The time-evolved trace distance for the case of $T=11.13$ maintains the highest relaxation rate, while for the other three temperature cases, the long-time dynamics cross over to a slower relaxation regime $\propto e^{\text{Re}(\lambda_{2})t}$ (see End Matter for details).
Therefore, we observe the occurrence of the canonical quantum Mpemba effect, where the hotter state at $T=11.13$ is exceptionally speedup and reaches the steady state faster than the colder states at $T=7$ and $T=9$.

We further quantify the acceleration capability of the canonical quantum Mpemba effect by defining the effective velocity $\upsilon_{\rm eff}$, which is the ratio of the initial trace distance to the critical time, 
$\upsilon_{\rm eff}=\frac{D(\varrho_{\text{th}}[T],\varrho_{\text{SS}})}{t_{c}[T]}$.
The solid line in Fig.~\ref{Fig:Ttim}(d) shows $\upsilon_{\rm eff}$ as a function of $T$.
We find that $\upsilon_{\rm eff}$ decreases with increasing temperature, and reaches a minimum around the steady-state effective temperature $T_{\text{SS}}\approx 5.77$.
This can be understood as the initial trace distance is minimized at this temperature, while the critical time is still finite, i.e., $D(\varrho_{\text{th}}[T],\varrho_{\text{SS}})\to0$ and $t_{c}[T]\to \text{const.}$, leading to the slowest effective velocity.
After that, the effective velocity $\upsilon_{\rm eff}$ increases and exhibits a sharp peak around $T\approx 11.13$. 
Within this regime, the relaxation acceleration increase monotonically as a function of the temperature.
We refer to this as the Mpemba zone, highlighted by the gray-shaded area in Fig.~\ref{Fig:Ttim}(d).

\begin{figure}[!tpb]
    \centering
    \includegraphics[width=0.475\textwidth]{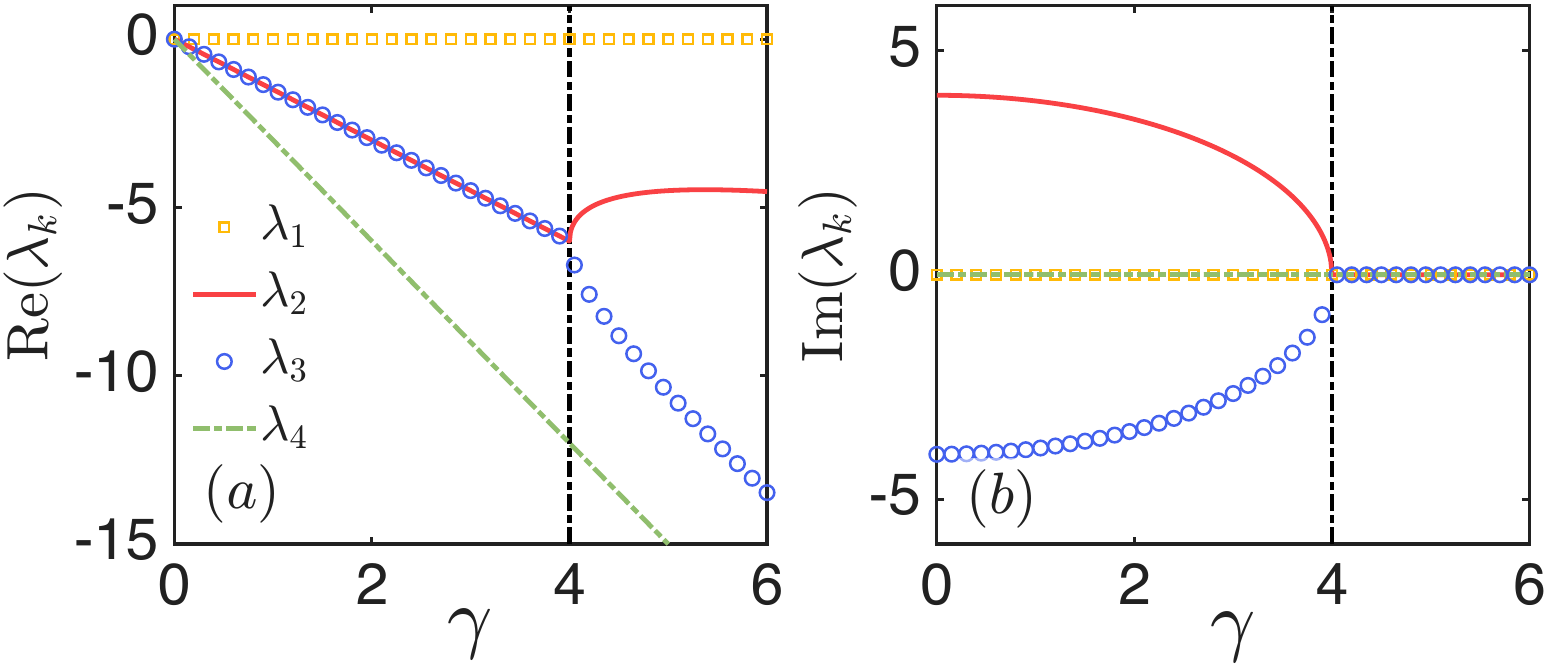}    
    \caption{\label{Fig:spectrum} The spectrum of Liouvillian $\Ls$. (a) real part and (b) imaginary part of the eigenvalues $\lambda_{k}$ as a function of the decay strength $\gamma$. The yellow squares, red solid, blue circles and green dashed line denote the eigenvalues $\lambda_{1}$, $\lambda_{2}$, $\lambda_{3}$, and $\lambda_{4}$, respectively. The other parameters are $\omega_{y}=0.01$ and $\omega_{z}=2$.}
\end{figure}

{\it Liouvillian exceptional point.}--- The exponentially accelerated relaxation dynamics in the quantum Mpemba effect is intrinsically linked to the presence of Liouvillian exceptional point (LEP)~\cite{Zhou2023b,Chatterjee2024,Zhang2025}.
In this work, the non-zero eigenvalues of $\Ls$ are analytically expressed as:
\begin{equation}
 \lambda_{k} = -2\gamma+2\sqrt{\frac{-R}{3}}\cos{\left(\frac{1}{3}\arccos{(-\frac{3Q}{2|R|}\sqrt{\frac{-3}{R}})}-\frac{2\pi k}{3}\right)},
\end{equation}
where $Q=-6\gamma\omega_{y}^2+4\gamma\omega_{z}^2-\frac{3}{4}\gamma^3$, $ R=4(\omega_{y}^2+\omega_{z}^2)-\frac{7}{4}\gamma^2$, and $k=2,3,4$ [see Supplementary Material (SM)~\cite{Supplemental}~\nocite{Minganti2018,Ciccarello2022,Lacroix2025} for details].
Panel (a) and (b) of ~\ref{Fig:spectrum} show the real and imaginary parts of the eigenvalues $\lambda_{k}$ as a function of the dissipation $\gamma$, respectively. We observe that the LEP occurs near $\gamma\approx4$.
Before reaching the LEP, $\lambda_{2}=\lambda_{3}^{*}$, preventing the elimination of $c_{k}$ ($k=2,3$) for thermal initial states~\cite{Zhang2025,Supplemental}.
Only when $\lambda_{2}$ is real does a critical temperature exist that satisfies $c_{2}=0$, e.g., $\gamma = 5$.
We locate the LEP via the discriminant $\Delta=\left(\frac{Q}{2}\right)^2+\left(\frac{R}{3}\right)^3$, where $\Delta>0$ yields one real and two complex-conjugate eigenvalues, while $\Delta\leq0$ gives three real eigenvalues. 
The LEP is thus exactly at $\Delta=0$~\cite{Supplemental}.

\begin{figure}[!htpb]
    \centering
    \includegraphics[width=0.475\textwidth]{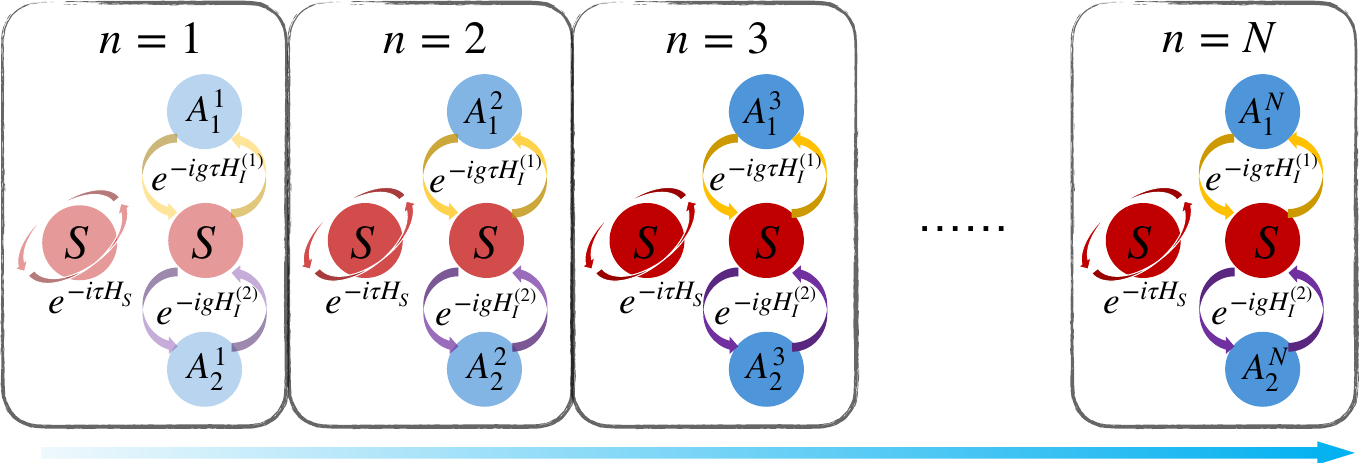}    
    \caption{\label{Fig:QCM} Schematic of the quantum collision model to realize Eq.~\eqref{Eq:ME}. 
    The system qubit $S$ interacts sequentially with two types of ancillas, $A_{1}$ and $A_{2}$, which are prepared in $|0\rangle$ state. 
    Each interaction is followed by a partial trace over the ancillas, simulating the effect of the environment.}
\end{figure}

{\it Experimental realization.}--- Equation~(\ref{Eq:ME}) can be readily implemented using currently available quantum simulation technologies combined with quantum collision models.
Quantum collision models have been widely used in simulations of open quantum systems~\cite{Cattaneo2021a,Landi2022a,Erbanni2023,Zhang2023h,Alves2024,Mahlow2024}, and have been experimentally realized~\cite{Cuevas2019,Garcia-Perez2020,Cattaneo2023,Cech2023,Li2025q}.
This approach provides a discrete-time framework for open quantum dynamics: the environment is simulated as a collection of ancillas, denoted by $\eta_{A}^{n}$, ($n=1,2,\dots$). 
The system's continuous interaction with its environment is discretized into a sequence of pairwise unitary interactions (``collisions") between the system and individual ancillas, after which the ancillas are discarded.
The system’s stroboscopic evolution is thus generated by composing these elementary collision steps.
Specifically, the system's reduced density matrix after the $(n+1)$th collision is obtained through the map: $\varrho^{n}_{S}\mapsto\varrho^{n+1}_{S}= \text{tr}_{\eta^{n+1}_{A}}[U_{I}U_{S}\varrho^{n}_{S}\otimes\eta^{n+1}_{A}U_{S}^{\dagger}U^{\dagger}_{I}]$, where $U_{S}=e^{-iH_{S}\tau}$ governs the free evolution, $U_{I}=e^{-igH_{I}\tau}$ describe the interactions, with $\tau$ denotes the duration of collision and $g$ is the interaction strength.
We simulate the two baths using two sets of ancillas, $\{A_{1}\}$ and $\{A_{2}\}$, as illustrated in Fig.~\ref{Fig:QCM}. 
The interaction Hamiltonian is
$H_{I}=H_{I}^{1}+H_{I}^{2}=\frac{1}{2}(\sigma^{x}_{S}\sigma^{x}_{A,1}+\sigma^{y}_{S}\sigma^{y}_{A,1})+\sigma^{y}_{S}\sigma^{x}_{A,2}$, with all ancillas initialized in $|0\rangle$ state, i.e., a zero-temperature state. 
In continuous-time limit ($\tau\to 0$), we recover the master equation [Eq.~\eqref{Eq:ME}], where $\gamma=g^2\tau$ (see SM~\cite{Supplemental} for details).
Each collision‐step unitary is then mapped onto a quantum circuit by decomposing it into qubit rotations, Hadamard, and CNOT gates,
\begin{equation}
\begin{aligned}
U_{S}=&R^{x}_{S}(\alpha)R^{z}_{S}(2\Omega\tau)R^{x}_{S}(-\alpha),\\
U_{I}^{1}=&R^{z}_{S}(\frac{\pi}{2})\text{H}_{A_{1}}\text{CNOT}_{A_{1},S}R^{y}_{S}(\theta)R^{y}_{A_{1}}(\theta)\text{CNOT}_{A_{1},\,S}R^{z}_{S}(-\frac{\pi}{2})\text{H}_{A_{1}},\\
U_{I}^{2}=&R^{x}_{S}(\frac{\pi}{2})\text{H}_{A_{2}}\text{CNOT}_{A_{2},\,S}R^{z}_{A_{2}}(2\theta)\text{CNOT}_{A_{2},\,S}R^{x}_{S}(-\frac{\pi}{2})\text{H}_{A_{2}},
\end{aligned}
\end{equation}
where $\Omega=\sqrt{\omega_{y}^2+\omega_{z}^2}$, $\alpha=\arctan2(\omega_{y},\omega_{z})$, and $\theta=g\tau$.
The partial‐trace operation is implemented by resetting the ancillas after each collision, which allows them to be reused in subsequent steps and reduces the required hardware to a simple $\tikz[baseline=(a.base),every node/.style={inner sep=0.2pt},x=1.3em]
{
  \node (a) at (0,0) {$A_1$};
  \node (b) at (1,0) {$S$};
  \node (c) at (2,0) {$A_2$};
  \draw (a) -- (b) -- (c);
}$ chain [see Fig.~\ref{Fig:Schematic}(c)].
Since thermal (Gibbs) state initialization in a quantum circuit is generally nontrivial~\cite{Wang2021,Edo2024,Consiglio2024}, we bypass direct initialization and exploit the linearity of the master equation in $\varrho_{\text{th}}$, writing its evolution as $\varrho_{\text{th}}(t)=\sum_{j}P_{j}e^{t\Ls}[\varrho_{j}]$, where $\varrho_{\text{th}}=\sum_{j}P_{j}|\psi_{j}\rangle\langle\psi_{j}|=\sum_{j}P_{j}\varrho_{j}$, with $P_{j}$ is the probability of the pure state $|\psi_{j}\rangle$.
By incorporating classical processing, one can reconstruct the evolution of a given thermal state [see Fig.~\ref{Fig:Schematic}(c)].
We verify the feasibility of our hybrid classical-quantum algorithm on the state-of-the-art quantum device. 
Detailed discussion and experimental results can be found in SM~\cite{Supplemental}.

{\it Conclusions.}--- In summary, we have introduced the canonical quantum Mpemba effect and demonstrated its occurrence in a dissipative qubit system.
By defining the steady-state effective temperature, we showed that the relaxation dynamics for initial thermal states toward the steady state constitute a genuine cooling process.
Notably, within the Mpemba zone, the relaxation dynamics exhibit exponential acceleration as the initial temperature increases.
This work provides the quantum analogue of the classical Mpemba effect, highlighting the fundamental role of temperature in anomalous relaxation dynamics, and offers a practical experimental scheme for its realization using superconducting qubits.
We anticipate that our findings can be extended to open quantum many‐body systems, and thereby spur further studies of anomalous non‐equilibrium phenomena.

\begin{acknowledgments}
{\it Acknowledgments.}--- 
The authors thank Jiasen Jin, Wenlin Li, and Jinghui Pi for reviewing the manuscript.
This work is supported by the Quantum Science and Technology-National Science and Technology Major Project (2024ZD0300600).
We acknowledge financial support from the National Natural Science Foundation of China under Grant No. 92565105 and 12204395, Hong Kong RGC No. 14301425, No. 24308323, and No. C4050-23GF, the Space Application System of China Manned Space Program, and CUHK Direct Grant.
\end{acknowledgments}


\resetappendix

\section{End Matter}

\begin{figure}[!htpb]
    \centering
    \includegraphics[width=0.48\textwidth]{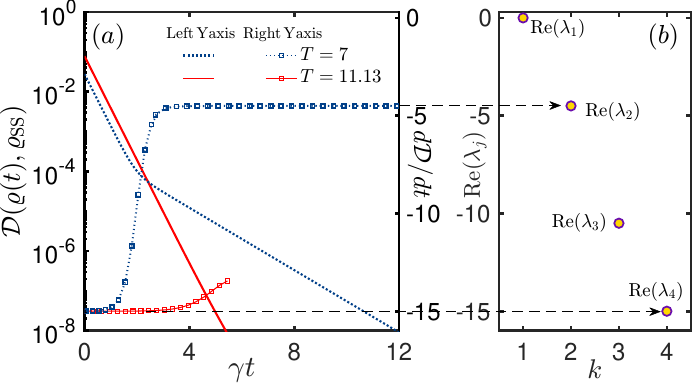}    
    \caption{\label{Fig:Liou} (a) Trace distance between the time-evolved state $\varrho(t)$ and the steady state $\varrho_{\text{SS}}$ (left axis, lines) and its derivative (right axis, markers connected by lines) are shown as functions of $\gamma t$. 
    Data are plotted for initial temperatures of $T=7$ (blue lines) and $T=11.13$ (red lines).
    (b) The distribution of the real part of the Liouvillian's eigenvalues $\text{Re}(\lambda_{k})$, ($k=1,2,3,4$). }
\end{figure}

{\it Appendix: The relaxation rate.}--- In Fig.~\ref{Fig:Ttim}(c), we show the time-evolved trace distance for different initial temperatures. 
We highlight that for the case of $T=7$, the relaxation proceeds in two stages: initially it relaxes at rate $\text{Re}(\lambda_{4})$, and then it abruptly switches to $\text{Re}(\lambda_{2})$ in the long-time dynamics.
For $T=11.13$, it follows a single-exponential form $\sim e^{\text{Re}(\lambda_{4})t}$.
By computing the time derivative of the trace distance, we extract numerical relaxation rates and directly compare them with the real parts of the corresponding
Liouvillian eigenvalues [Figs.~\ref{Fig:Liou}(a)–(b)], revealing excellent agreement.

\bibliography{filterref}
\end{document}


\beginsupplement

\title{Supplementary Material for ``Canonical Quantum Mpemba Effect in a Dissipative Qubit"}

\author{Xingli Li\orcidlink{0000-0003-2339-6557}}
\affiliation{Department of Physics, The Chinese University of Hong Kong, Shatin, New Territories, Hong Kong, China}

\author{Yan Li\orcidlink{0009-0007-7218-133X}}
\affiliation{Department of Physics, The Chinese University of Hong Kong, Shatin, New Territories, Hong Kong, China}

\author{Yangqian Yan\orcidlink{0000-0002-3237-5945}}
\email{yqyan@cuhk.edu.hk}
\affiliation{Department of Physics, The Chinese University of Hong Kong, Shatin, New Territories, Hong Kong, China}
\affiliation{State Key Laboratory of Quantum Information Technologies and Materials, The Chinese University of Hong Kong, Hong Kong SAR, China}
\affiliation{The Chinese University of Hong Kong Shenzhen Research Institute, 518057 Shenzhen, China
}

\begin{abstract}

\end{abstract}
\date{\today}
\maketitle
\tableofcontents

\section{Analytic solution of the Liouvillian}
In this section, we derive the analytical expression of the Liouvillian's eigenvalues and eigenvectors for the dissipative qubit system discussed in the main text. 
Based on those analytical results, we determine the location of the Liouvillian exceptional point in the parameter space and obtain the critical temperature at which the overlap between the initial thermal state and the Liouvillian eigenmode of the slowset mode becomes zero.

The Liouvillian $\Ls$ can be represented as a linear operator in Liouville space by applying the vectorization to the operators in Hilbert space, e.g., $\varrho=\sum_{j,k=1}^{N}\varrho_{jk}|j\rangle\langle k|\mapsto|\varrho\rangle\rangle=\sum_{j,k=1}^{N}\varrho_{jk}|j\rangle\otimes|k\rangle$.
Therefore, the matrix representation of the Liouvillian is~\cite{Minganti2018}
\begin{equation}
\Ls = -i[H\otimes\mathbb{I}-\mathbb{I}\otimes H^{\text{T}}]+\sum_{j}\frac{\gamma_{j}}{2}(2o_{j}\otimes o_{j}^{*}-o_{j}^{\dagger}o_{j}\otimes\mathbb{I}-\mathbb{I}\otimes o_{j}^{\text{T}}o_{j}^{*}),  
\end{equation}
where $o_{j}$ and $\gamma_{j}$ are the jump operator and decay rate for the $j$th dissipation channel, respectively, with the superscripts $\text{T}$ and $*$ denote the transpose and conjugate operations, and $\mathbb{I}$ is the identity operator.
In this work, the Liouvillian is specifically given as 
\begin{equation}
\frac{d}{dt}
\begin{pmatrix}
\varrho_{00}\\  
\varrho_{01}\\  
\varrho_{10}\\  
\varrho_{11}\\  
\end{pmatrix}
=\Ls
\begin{pmatrix}
\varrho_{00}\\  
\varrho_{01}\\  
\varrho_{10}\\  
\varrho_{11}\\  
\end{pmatrix},
\qquad
\Ls = 
\begin{pmatrix}
-2\gamma & -\omega_{y} & - \omega_{y} & \gamma\\
\omega_{y} & -2i\omega_{z}-\frac{3\gamma}{2} & -\gamma & -\omega_{y}\\
\omega_{y} & -\gamma & 2i\omega_{z}-\frac{3\gamma}{2} & -\omega_{y}\\
2\gamma & \omega_{y} & \omega_{y} & -\gamma\\
\end{pmatrix}.
\end{equation}
Since the density matrix preserves the trace, i.e., $\text{Tr}[\varrho_{\text{SS}}]=\varrho_{00}+\varrho_{11}=1$, the Liouvillian can be reduced into a $3\times3$ matrix by eliminating $\varrho_{11}$ via the constraint $\varrho_{11}=1-\varrho_{00}$. 
The reduced Liouvillian reads
\begin{equation}
\Ls_{\text{reduced}}\begin{pmatrix}
\varrho_{00}\\
\varrho_{01}\\
\varrho_{10}
\end{pmatrix}
=A\begin{pmatrix}
\varrho_{00}\\
\varrho_{01}\\
\varrho_{10}
\end{pmatrix}+B=
\begin{pmatrix}
-3\gamma & -\omega_{y} & -\omega_{y}\\
2\omega_{y} &  -2i\omega_{z}-\frac{3\gamma}{2} & -\gamma\\
2\omega_{y} &  -\gamma &  2i\omega_{z}-\frac{3\gamma}{2}
\end{pmatrix}\begin{pmatrix}
\varrho_{00}\\
\varrho_{01}\\
\varrho_{10}
\end{pmatrix}+
\begin{pmatrix}
\gamma\\
-\omega_{y}\\
-\omega_{y}\\ 
\end{pmatrix},
\end{equation}
The non-zero eigenvalues of the Liouvillian are determined by solving the characteristic equation $\det(A-\lambda\mathbb{I})=0$, yielding:
\begin{equation}
\lambda^{3}+6\gamma\lambda^2+(4\omega_{y}^2+4\omega_{z}^2+\frac{41\gamma^2}{4})\lambda+(2\gamma\omega_{y}^2+12\gamma\omega_{z}^2+\frac{15\gamma^3}{4})=0,  
\label{Eq:Cubic}
\end{equation}
Applying the Cardano formula for cubic equations, we obtain the eigenvalues
\begin{equation}
 \lambda_{k} = -2\gamma+2\sqrt{\frac{-R}{3}}\cos{\left(\frac{1}{3}\arccos{(-\frac{3Q}{2|R|}\sqrt{\frac{-3}{R}})}-\frac{2\pi k}{3}\right)},
 \label{Eq:EigenFun}
\end{equation}
for $k=2,3,4$, with the parameters:
\begin{equation}
Q=-6\gamma\omega_{y}^2+4\gamma\omega_{z}^2-\frac{3}{4}\gamma^3, \quad R=4(\omega_{y}^2+\omega_{z}^2)-\frac{7}{4}\gamma^2,
\end{equation}
It should be emphasized that the eigenvalues $\lambda_{k}$ are not ordered according to their real parts at this stage. 
Thus, the subscript $k$ does not indicate the ordering.

\begin{figure}[!tpb]
    \centering
    \includegraphics[width=1\textwidth]{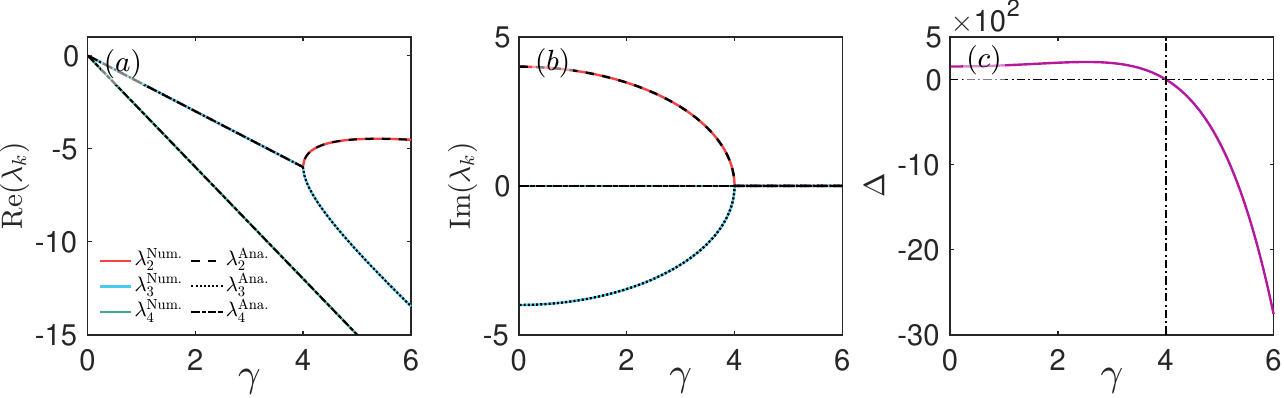}    
    \caption{\label{Fig:Lsupp} (a)-(b) Comparison of numerical (denoted by `Num.',) and analytical (denoted by `Ana.') results of the Liouvillian's eigenvalues as a function of the decay rate $\gamma$: (a) real part and (b) imaginary part. 
    (c) The discriminant $\Delta$ changes with the decay rate.
    The black dashdotted lines indicates where $\Delta=0$.
    The parameters are $\omega_{y}=0.01$ and $\omega_{z}=2$.}
\end{figure}

Figures~\ref{Fig:Lsupp}(a) and (b) show the real and imaginary parts of the Liouvillian's eigenvalues, obtained numerically (colored solid lines) and analytically (black dashed, dotted, dashdotted lines), as a function of the decay rate, respectively. 
The analytical and numerical results show excellent agreement for both the real and imaginary parts, with
Eq.~\eqref{Eq:EigenFun} accurately capturing all the non-zero eigenvalues.
Moreover, the discriminant $\Delta=\left(\frac{Q}{2}\right)^2+\left(\frac{R}{3}\right)^3$ determines the properties of the roots of Eq.~\eqref{Eq:Cubic}, i.e.,
\begin{enumerate}
    \item $\Delta > 0$: One real root and two non-real complex conjugate roots. 
    \item $\Delta = 0$: Three real roots (counting multiplicity):
    \begin{itemize}
        \item $\left(\dfrac{Q}{2}\right)^2 = -\left(\dfrac{R}{3}\right)^3 = 0$: Triple real root.
        \item $\left(\dfrac{Q}{2}\right)^2 = -\left(\dfrac{R}{3}\right)^3 \neq 0$: One distinct real root and one double real root.
    \end{itemize}
    \item $\Delta < 0$: Three distinct real roots.
\end{enumerate}
This implies that when $\Delta=0$, the eigenvalues of the Liouvillian transition from having complex roots to being entirely real.
In Fig.~\ref{Fig:Lsupp}(c), the numerics shows that when $\Delta\approx 0$, the decay rate approaches $\gamma=4$, consistent with the transition point shown in Figs.~\ref{Fig:Lsupp}(a) and (b).

Therefore, the Liouvillian exceptional point emerges in parameter space at $\Delta=0$, which leads to the following equation
\begin{equation}
25x^3+60(\omega_{z}^2-26\omega_{y}^2)x^2+48(164\omega_{y}^2\omega_{z}^2-53\omega_{y}^2-8\omega_{z}^2)x-1024(\omega_{y}^2+\omega_{z}^2)^3=0, 
\end{equation}
where $x=\gamma^2$. 
We still solve the cubic equation using the Cardano formula, defining the coefficients $a=25$, $b=60(\omega_{z}^2-26\omega_{y}^2)$, $c=48(164\omega_{y}^2\omega_{z}^2-53\omega_{y}^2-8\omega_{z}^2)$
and $d=-1024(\omega_{y}^2+\omega_{z}^2)^3$.
The roots can be expressed as 
\begin{equation}
x_{k}= -\frac{b}{3a}+2\sqrt{\frac{-r}{3}}\cos{\left(\frac{1}{3}\arccos{(\frac{3q}{2|r|}\sqrt{\frac{-3}{r}})}-\frac{2\pi k}{3}\right)},
\end{equation}
where $k=1,2,3$, $q=\frac{bc}{a^2}-\frac{2b^3}{27a^3}-\frac{d}{a}$ and $r=\frac{c}{a}-\frac{b^2}{3a^2}$. 
The critical decay rate is obtained by $\gamma_{c}=\sqrt{x_{k}}$, and we have to select the physically solution where $\gamma$ is real and non-negative.
For $\omega_{y}=0.01$ and  $\omega_{z}=2$, the computed critical decay rate $\gamma_{c}\approx 4$ matches the numerical result.

Correspondingly, the left and right eigenvectors of the Liouvillian associated with non-zero eigenvalues read
\begin{equation}
l_{k} = \frac{1}{n_{k}^{l}}
\begin{pmatrix}
\gamma-\lambda_{k}&
-\frac{\lambda_{k}^2+3\gamma\lambda_{k}}{\omega_{y}}(1+\frac{4i\omega_{z}}{\gamma+2\lambda_{k}})\\    
-\frac{\lambda_{k}^2+3\gamma\lambda_{k}}{\omega_{y}}(1-\frac{4i\omega_{z}}{\gamma+2\lambda_{k}})&
\gamma+\lambda_{k}
\end{pmatrix},\qquad
r_{k} = \frac{1}{n_{k}^{r}}
\begin{pmatrix}
2\omega_{y}(\lambda_{k}+\frac{\gamma}{2})&
(3\gamma+\lambda_{k})(2i\omega_{z}-\lambda_{k}-\frac{\gamma}{2}) \\
-(3\gamma+\lambda_{k})(2i\omega_{z}+\lambda_{k}+\frac{\gamma}{2}) &
-2\omega_{y}(\lambda_{k}+\frac{\gamma}{2})    
\end{pmatrix},
\end{equation}
where $\lambda_{k}$ is the corresponding eigenvalue, $n_{k}^{l}$ and $n_{k}^{r}$ denote normalization constants for the left and right eigenmatrices, respectively.
The preceding analysis addressed the Liouvillian's eigenvectors with non-zero eigenvalues. For the steady state, given by the eigenvector with zero eigenvalue, we obtain an analytical expression by solving the Bloch equation, which reads
\begin{equation}
\begin{aligned}
\frac{d}{dt}\langle\sigma^{x}_{S}\rangle=&2\omega_{y}\langle\sigma^{z}_{S}\rangle-2\omega_{z}\langle\sigma^{y}_{S}\rangle-\frac{5\gamma}{2}\langle\sigma^{x}_{S}\rangle,\\
\frac{d}{dt}\langle\sigma^{y}_{S}\rangle=&2\omega_{z}\langle\sigma^{x}_{S}\rangle-\frac{\gamma}{2}\langle\sigma^{y}_{S}\rangle,\\
\frac{d}{dt}\langle\sigma^{z}_{S}\rangle=&-2\omega_{y}\langle\sigma^{x}_{S}\rangle-3\gamma\langle\sigma^{z}_{S}\rangle-\gamma.
\end{aligned}   
\end{equation}
Requiring $d\langle\sigma^{x}_{S}\rangle/dt=d\langle\sigma^{y}_{S}\rangle/dt=d\langle\sigma^{z}_{S}\rangle/dt=0$, we then obtain the steady-state solution,
\begin{equation}
\langle\sigma^{x}_{S}\rangle_{\text{SS}}=  -\frac{4\omega_{y}\gamma}{8\omega^{2}_{y}+48\omega^{2}_{z}+15\gamma^2}, \quad 
\langle\sigma^{y}_{S}\rangle_{\text{SS}}=  -\frac{16\omega_{y}\omega_{z}}{8\omega^{2}_{y}+48\omega^{2}_{z}+15\gamma^2},\quad 
\langle\sigma^{z}_{S}\rangle_{\text{SS}}=  -\frac{16\omega_{z}^2+5\gamma^2}{8\omega^{2}_{y}+48\omega^{2}_{z}+15\gamma^2}.
\end{equation}
Hence, the steady state can be reconstructed by $\varrho_{\text{SS}}=\frac{1}{2}(\mathbb{I}+\langle\sigma^{x}_{S}\rangle_{\text{SS}}\cdot\sigma^{x}_{S}+\langle\sigma^{y}_{S}\rangle_{\text{SS}}\cdot\sigma^{y}_{S}+\langle\sigma^{z}_{S}\rangle_{\text{SS}}\cdot\sigma^{z}_{S})$.

With the complete set of left and right eigenvectors of the Liouvillian at hand, we can then analytically compute the overlap between a thermal state and these eigenvectors. 
This analysis allows us to identify at which initial temperature the thermal state can exhibit exponential acceleration of relaxation.
We begin by presenting the analytical expression for the thermal state, which can be expressed in the following way
\begin{equation}
\varrho_{\text{th}}[T]=
\begin{pmatrix}
\frac{1}{2}-\beta_{\text{eff}}\Omega_{z}&i\beta_{\text{eff}}\Omega_{y}, \\
-i\beta_{\text{eff}}\Omega_{y}&
\frac{1}{2}+\beta_{\text{eff}}\Omega_{z}
\end{pmatrix},
\end{equation}
where $\beta_{\text{eff}}=\tanh\left(\frac{\sqrt{\omega_{y}^2+\omega_{z}^2}}{T}\right)$ and  $\Omega_{y(z)}=\frac{\omega_{y(z)}}{2\sqrt{\omega_{y}^2+\omega_{z}^2}}$.
Then, the overlap $c_{k}=|\text{Tr}[l_{k}\varrho_{\text{th}}[T]]|$ is given by
\begin{equation}
c_{k}=\left|\frac{f(\lambda_{k})}{n_{k}^{l}}\right|,
\end{equation}
with $f(\lambda_{k})=\gamma-2\beta_{\text{eff}}\Omega_{z}\lambda_{k}\frac{5\gamma}{\gamma+2\lambda_{k}}$, where $\lambda_{k}$ is the eigenvalue corresponding to the eigenvector $r_{k}$ as defined previously.
To achieve the exponentially accelerated relaxation, we require $c_{2}=0$, leading to the condition $f(\lambda_{2})=0$. 
This demands that the eigenvalue of the slowest dissipative mode at this instant satisfies:
\begin{equation}
\lambda_{2}=-\frac{\gamma}{2-10\Omega_{z}\beta_{\text{eff}}}.   
\end{equation}
However, the parameters are real-valued, i.e., $(\gamma,\Omega_{z},\beta_{\text{eff}})\in\mathbb{Re}$, and thus $f(\lambda_{2})=0$ admits a solution only if $\lambda_{2}\in\mathbb{Re}$.
This necessity for a real $\lambda_{2}$ explains why the exponential acceleration emerges only after the system goes through the Liouvillian exceptional point. Specifically, it is the crossing of the Liouvillian EP that drives the coalescence and subsequent splitting of the eigenvalues, causing the originally complex to become real and thereby satisfying the critical solvability condition.
Additionally, once $\lambda_{2}$ satisfies the real-valued condition, the critical temperature $T_{c}$ can be determined,
\begin{equation}
T_{c}=\frac{\sqrt{\omega_{y}^2 + \omega_{y}^2}}{\arctan\frac{2\lambda_{2}+\gamma}{10\lambda_{2}\Omega_{z}}}.
\end{equation}
Under the parameter condition $\omega_{y}=0.01$, $\omega_{z}=2$, and $\gamma=5$, we obtain $\lambda_{2}=-4.5$ and $T_{c}\approx 11.1305$, in perfect agreement with our numerical results presented in the main text.

\section{The derivation of the quantum master equation in the quantum collision model framework}

In this section, we introduce how to unravel the quantum master equation in the quantum collision model framework, with the goal of mapping it onto the quantum circuits.
In quantum collision model framework, the environment is composed of a set of ancillas that interact with the system sequentially, and the interaction between the system and the ancillas is described by a unitary evolution. 
The sequential interaction can be described by the following dynamics mapping
\begin{equation}
  \varrho^{n}_{S}\mapsto \varrho^{n+1}_{S}=\text{Tr}_{\eta^{n+1}_{E}}[U_{I}U_{S}(\varrho^{n}_{S}\otimes\eta^{n+1}_{E})U_{S}^{\dagger}U_{I}^{\dagger}],
  \label{Eq:QCMmap}
\end{equation}
where $\varrho_{S}^{n}$ is the state of the system after the $n$th interaction, $\eta^{n+1}_{E}$ is the $(n+1)$th ancilla, $U_{S}$ is the self-evolution of the system, which is given by $H_{S}=e^{-iH_{S}\tau}$, and $U_{I}=e^{-igH_{I}\tau}$ describes the interaction between the system and the ancillas, where $g$ is the coupling strength and $\tau$ is the interaction time. 
Moreover, Eq.~\eqref{Eq:QCMmap} describes a Markovian process, which is due to the fact that the system interacts with a fresh (i.e., uncorrelated) ancilla at each step~\cite{Ciccarello2022,Lacroix2025}.

After establishing the foundation of the quantum collision model, we focus on the specific system under investigation in this work.
The system Hamiltonian reads
\begin{equation}
    H_{S}=\omega_{y}\sigma^{y}_{S}+\omega_{z}\sigma^{z}_{S}.
\end{equation}
We consider two sets of identical ancillas, i.e., $\{A_{1}\}$ and $\{A_{2}\}$, to simulate the two baths in the quantum master equation.
The free Hamiltonian of the ancillas is given by $H_{A,j}=\sigma_{A,j}^{z}$,
which the corresponding thermal state is $\eta_{A,j}=e^{-\beta_{j}\sigma^{z}_{A,j}}/\text{Tr}[e^{-\beta_{j}\sigma^{z}_{A,j}}]$, where $\beta_{j}=1/T_{j}$ is the inverse temperature of the $j$th bath with $j=1,2$.
Then, the interaction Hamiltonian between the system and the ancillas reads
\begin{equation}
 H_{I} = \sigma^{-}_{S}\sigma^{+}_{A,1}+\sigma^{y}_{S}\sigma^{+}_{A,2} + \sigma^{+}_{S}\sigma^{-}_{A,1}+\sigma^{y}_{S}\sigma^{-}_{A,2},
\end{equation}
where $\sigma^{\pm}_{\alpha}=(\sigma^{x}_{\alpha}\pm i\sigma^{y}_{\alpha})/2$ is the raising and lowering operators of the system and the two ancillas, respectively.
Therefore, the dynamics mapping~\eqref{Eq:QCMmap} can be written as
\begin{equation}
\varrho^{n+1}_{S}=\text{Tr}_{\eta^{n+1}_{A,1},\eta^{n+1}_{A,2}}[U_{I}U_{S}(\varrho\otimes\eta_{A,1}^{n+1}\otimes\eta_{A,2}^{n+1})U_{S}^{\dagger}U_{I}^{\dagger}],
\label{Eq:QCMmap2}
\end{equation}
where $\eta_{A,1}^{n+1}$ and $\eta_{A,2}^{n+1}$ are the $(n+1)$th ancillas of the two baths, respectively.
Since the interaction time $\tau$ is small, we can expand the unitary operators $U_{S}$ and $U_{I}$ to the second order in $\tau$ by using the Taylor expansion, i.e.,
\begin{equation}
U_{S}=\mathbb{I}-i\tau H_{S}-\frac{\tau^2}{2}H_{S}^2+o(\tau^3),
\end{equation}
and 
\begin{equation}
U_{I}=\mathbb{I}-ig\tau H_{I}-\frac{\tau^2g^2}{2}H_{I}^2+o(\tau^3).
\end{equation}
The joint unitary operator $U_{I}U_{S}$ then can then be approximated as
\begin{equation}
U_{I}U_{S}\approx(\mathbb{I}-ig\tau H_{I}-i\tau H_{S}-\frac{\tau^2g^2}{2}H_{I}^2-\frac{\tau^2}{2}H_{S}^2).    
\end{equation}
Equation~\eqref{Eq:QCMmap2} can then be approximated as
\begin{equation}
\begin{aligned}
\varrho^{n+1}_{S}=&\text{Tr}_{\eta^{n+1}_{A,1},\eta^{n+1}_{A,2}}[U_{I}U_{S}(\varrho^{n}_{S}\otimes\eta_{A,1}^{n+1}\otimes\eta_{A,2}^{n+1})U_{S}^{\dagger}U_{I}^{\dagger}]\\
\approx&\text{Tr}_{\eta^{n+1}_{E,1},\eta^{n+1}_{E,2}}[(\mathbb{I}-ig\tau H_{I}-i\tau H_{S}-\frac{\tau^2g^2}{2}H_{I}^2-\frac{\tau^2}{2}H_{S}^2)(\varrho^{n}_{S}\otimes\eta_{E,1}^{n+1}\otimes\eta_{E,2}^{n+1})(\mathbb{I}+ig\tau H_{I}+i\tau H_{S}-\frac{\tau^2g^2}{2}H_{I}^2-\frac{\tau^2}{2}H_{S}^2)]\\
=&\text{Tr}_{\eta^{n+1}_{A,1},\eta^{n+1}_{A,2}}[(\varrho^{n}_{S}\otimes\eta_{A,1}^{n+1}\otimes\eta_{A,2}^{n+1})]+ig\tau\text{Tr}_{\eta^{n+1}_{A,1},\eta^{n+1}_{A,2}}[(\varrho^{n}_{S}\otimes\eta_{A,1}^{n+1}\otimes\eta_{A,2}^{n+1})H_{I}]+i\tau\text{Tr}_{\eta^{n+1}_{A,1},\eta^{n+1}_{A,2}}[(\varrho^{n}_{S}\otimes\eta_{A,1}^{n+1}\otimes\eta_{A,2}^{n+1})H_{S}]\\
&-\frac{\tau^2g^2}{2}\text{Tr}_{\eta^{n+1}_{A,1},\eta^{n+1}_{A,2}}[(\varrho^{n}_{S}\otimes\eta_{A,1}^{n+1}\otimes\eta_{A,2}^{n+1})H_{I}^{2}]-\frac{\tau^2}{2}\text{Tr}_{\eta^{n+1}_{A,1},\eta^{n+1}_{A,2}}[(\varrho^{n}_{S}\otimes\eta_{A,1}^{n+1}\otimes\eta_{A,2}^{n+1})H_{S}^{2}]-ig\tau\text{Tr}_{\eta^{n+1}_{A,1},\eta^{n+1}_{A,2}}[H_{I}(\varrho^{n}_{S}\otimes\eta_{A,1}^{n+1}\otimes\eta_{A,2}^{n+1})]\\
&+g^2\tau^2\text{Tr}_{\eta^{n+1}_{A,1},\eta^{n+1}_{A,2}}[H_{I}(\varrho^{n}_{S}\otimes\eta_{A,1}^{n+1}\otimes\eta_{A,2}^{n+1})H_{I}]+g\tau^2\text{Tr}_{\eta^{n+1}_{A,1},\eta^{n+1}_{A,2}}[H_{I}(\varrho^{n}_{S}\otimes\eta_{A,1}^{n+1}\otimes\eta_{A,2}^{n+1})H_{S}]+\frac{ig^3\tau^3}{2}\text{Tr}_{\eta^{n+1}_{A,1},\eta^{n+1}_{A,2}}[H_{I}(\varrho^{n}_{S}\otimes\eta_{A,1}^{n+1}\otimes\eta_{A,2}^{n+1})H_{I}^2]\\
&+\frac{ig\tau^3}{2}\text{Tr}_{\eta^{n+1}_{A,1},\eta^{n+1}_{E,2}}[H_{I}(\varrho^{n}_{S}\otimes\eta_{A,1}^{n+1}\otimes\eta_{A,2}^{n+1})H_{S}^2]-i\tau\text{Tr}_{\eta^{n+1}_{A,1},\eta^{n+1}_{A,2}}[H_{S}(\varrho^{n}_{S}\otimes\eta_{A,1}^{n+1}\otimes\eta_{A,2}^{n+1})]+g\tau^{2}\text{Tr}_{\eta^{n+1}_{A,1},\eta^{n+1}_{A,2}}[H_{S}(\varrho^{n}_{S}\otimes\eta_{A,1}^{n+1}\otimes\eta_{A,2}^{n+1})H_{I}]\\
&+\tau^2\text{Tr}_{\eta^{n+1}_{A,1},\eta^{n+1}_{A,2}}[H_{S}(\varrho^{n}_{S}\otimes\eta_{A,1}^{n+1}\otimes\eta_{A,2}^{n+1})H_{S}]+\frac{i\tau^{3}g^2}{2}\text{Tr}_{\eta^{n+1}_{A,1},\eta^{n+1}_{A,2}}[H_{S}(\varrho^{n}_{S}\otimes\eta_{A,1}^{n+1}\otimes\eta_{A,2}^{n+1})H^{2}_{I}]+\frac{i\tau^3}{2}\text{Tr}_{\eta^{n+1}_{A,1},\eta^{n+1}_{A,2}}[H_{S}(\varrho^{n}_{S}\otimes\eta_{A,1}^{n+1}\otimes\eta_{A,2}^{n+1})H^{2}_{S}]\\
&-\frac{\tau^2g^2}{2}\text{Tr}_{\eta^{n+1}_{A,1},\eta^{n+1}_{A,2}}[H_{I}^2(\varrho^{n}_{S}\otimes\eta_{A,1}^{n+1}\otimes\eta_{A,2}^{n+1})]-\frac{i\tau^3g^3}{2}\text{Tr}_{\eta^{n+1}_{A,1},\eta^{n+1}_{A,2}}[H_{I}^2(\varrho^{n}_{S}\otimes\eta_{A,1}^{n+1}\otimes\eta_{A,2}^{n+1})H_{I}]-\frac{i\tau^3g^2}{2}\text{Tr}_{\eta^{n+1}_{A,1},\eta^{n+1}_{A,2}}[H_{I}^2(\varrho^{n}_{S}\otimes\eta_{A,1}^{n+1}\otimes\eta_{A,2}^{n+1})H_{S}]\\
&+\frac{\tau^4g^4}{2}\text{Tr}_{\eta^{n+1}_{A,1},\eta^{n+1}_{A,2}}[H_{I}^2(\varrho^{n}_{S}\otimes\eta_{A,1}^{n+1}\otimes\eta_{A,2}^{n+1})H^{2}_{I}]+\frac{\tau^4g^2}{2}\text{Tr}_{\eta^{n+1}_{A,1},\eta^{n+1}_{A,2}}[H_{I}^2(\varrho^{n}_{S}\otimes\eta_{A,1}^{n+1}\otimes\eta_{A,2}^{n+1})H^{2}_{S}]-\frac{\tau^2}{2}\text{Tr}_{\eta^{n+1}_{A,1},\eta^{n+1}_{A,2}}[H_{S}^2(\varrho^{n}_{S}\otimes\eta_{A,1}^{n+1}\otimes\eta_{A,2}^{n+1})]\\
&-\frac{i\tau^3g}{2}\text{Tr}_{\eta^{n+1}_{A,1},\eta^{n+1}_{A,2}}[H_{S}^2(\varrho^{n}_{S}\otimes\eta_{A,1}^{n+1}\otimes\eta_{A,2}^{n+1})H_{I}]-\frac{i\tau^3}{2}\text{Tr}_{\eta^{n+1}_{A,1},\eta^{n+1}_{A,2}}[H_{S}^2(\varrho^{n}_{S}\otimes\eta_{A,1}^{n+1}\otimes\eta_{A,2}^{n+1})H_{S}]+\frac{\tau^4g^2}{2}\text{Tr}_{\eta^{n+1}_{A,1},\eta^{n+1}_{A,2}}[H_{S}^2(\varrho^{n}_{S}\otimes\eta_{A,1}^{n+1}\otimes\eta_{A,2}^{n+1})H^{2}_{I}]\\
&+\frac{\tau^4}{2}\text{Tr}_{\eta^{n+1}_{A,1},\eta^{n+1}_{A,2}}[H_{S}^2(\varrho^{n}_{S}\otimes\eta_{A,1}^{n+1}\otimes\eta_{A,2}^{n+1})H^{2}_{S}].
\end{aligned}
\end{equation}
Recalling the continuous-time limit $(\tau\to0)$ and assuming that $g^2\tau=\text{const.}$, we ignore the higher-order terms. 
Specifically, we ignore the terms of the form $g^{m}\cdot\tau^{n}$ with satisfying $m<2(n-1)$, and we obtain the following equation
\begin{equation}
\begin{aligned}  
 \lim_{\tau\to0}\frac{\varrho^{n+1}_{S}-\varrho^{n}_{S}}{\tau}\approx &-i[H_{S},\varrho^{n}_{S}]+ g^2\tau\text{Tr}_{\eta^{n+1}_{A,1},\eta^{n+1}_{A,2}}[H_{I}(\varrho^{n}_{S}\otimes\eta_{A,1}^{n+1}\otimes\eta_{A,2}^{n+1})H_{I}-(\varrho^{n}_{S}\otimes\eta_{A,1}^{n+1}\otimes\eta_{A,2}^{n+1})\frac{H_{I}^{2}}{2}-\frac{H_{I}^{2}}{2}(\varrho^{n}_{S}\otimes\eta_{A,1}^{n+1}\otimes\eta_{A,2}^{n+1})].
\end{aligned}
\end{equation}
We consider the temperature of the two baths is zero, i.e., $\eta_{A,j}=e^{-\beta_{j}\sigma^{z}_{A,j}}/\text{Tr}[e^{-\beta_{j}\sigma^{z}_{A,j}}]$ with $\beta_{j}\to\infty$ for $j=1,2$. 
Therefore, we initialize the ancillas in the state 
$|\downarrow\rangle\langle\downarrow|$, i.e., $\eta_{A,1}^{n+1}=\eta_{A,2}^{n+1}=|\downarrow\rangle\langle\downarrow|$, then we have 
\begin{equation}
\begin{aligned}
\text{Tr}_{\eta^{n+1}_{A,1},\eta^{n+1}_{A,2}}[H_{I}(\varrho^{n}_{S}\otimes\eta_{A,1}^{n+1}\otimes\eta_{A,2}^{n+1})H_{I}]=& \sigma^{-}_{S}\varrho^{n}_{S}\sigma^{+}_{S}+\sigma^{y}_{S}\varrho^{n}_{S}\sigma^{y}_{S},\\
\text{Tr}_{\eta^{n+1}_{A,1},\eta^{n+1}_{A,2}}[(\varrho^{n}_{S}\otimes\eta_{A,1}^{n+1}\otimes\eta_{A,2}^{n+1})\frac{H_{I}^{2}}{2}]=&\varrho^{n}_{S}\sigma^{+}_{S}\sigma^{-}_{S}+\varrho^{n}_{S}\sigma^{y}_{S}\sigma^{y}_{S},\\
\text{Tr}_{\eta^{n+1}_{A,1},\eta^{n+1}_{A,2}}[\frac{H_{I}^{2}}{2}(\varrho^{n}_{S}\otimes\eta_{A,1}^{n+1}\otimes\eta_{A,2}^{n+1})]=&\sigma^{+}_{S}\sigma^{-}_{S}\varrho^{n}_{S}+\sigma^{y}_{S}\sigma^{y}_{S}\varrho^{n}_{S},
\end{aligned}
\end{equation}
which leads to the following equation
\begin{equation}
\lim_{\tau\to0}\frac{\varrho^{n+1}_{S}-\varrho^{n}_{S}}{\tau}\approx -i[H_{S},\varrho^{n}_{S}]+ \frac{g^2\tau}{2}(2\sigma^{-}_{S}\varrho^{n}_{S}\sigma^{+}_{S}-\left\{\varrho^{n}_{S},\sigma^{+}_{S}\sigma^{-}_{S}\right\})+\frac{g^2\tau}{2}(2\sigma^{y}_{S}\varrho^{n}_{S}\sigma^{y}_{S}-\left\{\varrho^{n}_{S},\sigma^{y}_{S}\sigma^{y}_{S}\right\}),
\label{Eq:QME}
\end{equation}
where $\{\cdot\}$ denotes the anti-commutator. 
Equation~\eqref{Eq:QME} is the quantum master equation in the main text with $\gamma_{-}=\gamma_{y}=g\tau^2$, which describes the dynamics of the system under the influence of two baths. 
Besides, the case of $\gamma_{-}\neq\gamma_{y}$ can be achieved by adjusting the coupling strength $g$ for the two baths.

\section{Experimental realization based on quantum circuits}

Building on the quantum collision model framework, in which the Lindblad master equation is decomposed into successive system–ancillas unitary interaction, we now map this stroboscopic dynamics into a quantum-circuit implementation.

First, the system’s free‐evolution unitary operator $U_{S}$ over time $\tau$ can be realized by the following single‐qubit gates,
\begin{equation}
U_{S}(\tau)=R^{x}_{S}(\alpha)R^{z}_{S}(2\Omega\tau)R^{x}_{S}(-\alpha),
\end{equation}
where $R^{x(z)}_{S}$ denotes the rotation gate about the $x(z)$-axis, with $\Omega=\sqrt{\omega_{y}^2+\omega_{z}^2}$ and $\alpha=\arctan2(\omega_{y},\omega_{z})$.
Here, we decompose $U_{S}$ into the basic rotation gates for demonstration.
In Qiskit, the single‐qubit rotation can be implemented directly by the $\texttt{U3}$ gate, which realizes an arbitrary rotation on the Bloch sphere via three Euler angles.

As for the system–ancillas interaction, because the system interacts simultaneously with two ancillas, we have to decompose the evolution operator $U_{I}$ into two-body interactions.
Concretely, we partition the interaction Hamiltonian into two terms $H_{I}=H_{I}^{1}+H_{I}^{2}$, with $H_{I}^{1}=\frac{1}{2}(\sigma^{x}_{S}\sigma^{x}_{A,1}+\sigma^{y}_{S}\sigma^{y}_{A,1})$ and $H_{I}^{2}=\sigma^{y}_{S}\sigma^{x}_{A,2}$.
Then, the unitary operators corresponding to these two parts can be realized using single-qubit rotations, Hadamard gates, and CNOT gates as follows
\begin{equation}
\begin{aligned}
U_{I}^{1}(\tau)=&R^{z}_{S}(\frac{\pi}{2})\text{H}_{A_{1}}\text{CNOT}_{A_{1},S}R^{y}_{S}(\theta)R^{y}_{A_{1}}(\theta)\text{CNOT}_{A_{1},\,S}R^{z}_{S}(-\frac{\pi}{2})\text{H}_{A_{1}},\\
U_{I}^{2}(\tau)=&R^{x}_{S}(\frac{\pi}{2})\text{H}_{A_{2}}\text{CNOT}_{A_{2},\,S}R^{z}_{A_{2}}(2\theta)\text{CNOT}_{A_{2},\,S}R^{x}_{S}(-\frac{\pi}{2})\text{H}_{A_{2}},
\end{aligned}
\end{equation}
where the rotation angle is $\theta=g\tau$.

\begin{figure}[!hb]
    \centering
    \includegraphics[width=0.8\textwidth]{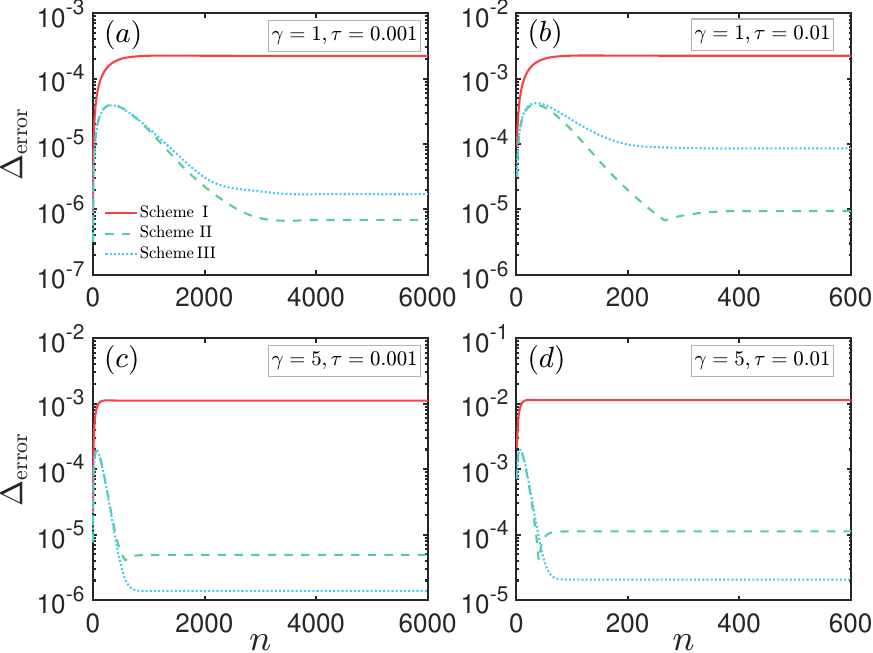}    
    \caption{\label{Fig:CMscheme} The error function as a function of the collision number $n$ under different parameters: (a) $\gamma=1,\tau=0.001$, (b) $\gamma=1,\tau=0.01$, (c) $\gamma=5,\tau=0.001$, and (d) $\gamma=5,\tau=0.01$. }
\end{figure}

To achieve the highest possible simulation accuracy, we systematically compare three implementation schemes based on the Suzuki-Trotter decomposition to realize the dynamical map in Eq.~\eqref{Eq:QCMmap}. 
\begin{equation}
   \varrho^{n+1}_{S}=\text{tr}_{\eta^{n+1}_{E}}[U_{\text{tot}}(\varrho^{n}_{S}\otimes\eta^{n+1}_{E})U^{\dagger}_{\text{tot}}],
\end{equation}
Scheme I (first-order Suzuki–Trotter),
\begin{equation}
U_{\text{tot}}=U_{I}^{1}(\tau)U_{I}^{2}(\tau)U_{S}(\tau),
\end{equation}
Scheme II (second-order Suzuki–Trotter),  
\begin{equation}
U_{\text{tot}}=U_{I}^{2}(\frac{\tau}{2})U_{I}^{1}(\tau)U_{I}^{2}(\frac{\tau}{2})U_{S}(\tau),
\end{equation}
Scheme III (variant second-order Suzuki–Trotter),
\begin{equation}
U_{\text{tot}}=U_{I}^{2}(\frac{\tau}{2})U_{S}(\frac{\tau}{3})U_{I}^{1}(\frac{\tau}{2})U_{S}(\frac{\tau}{3})U_{I}^{1}(\frac{\tau}{2})U_{S}(\frac{\tau}{3})U_{I}^{2}(\frac{\tau}{2}).
\end{equation}
The reason for examining the above three schemes lies in the fact that, for a fixed evolution time $t$, the choice of interaction time $\tau$ not only sets the approximation error of both the Taylor expansion and the Suzuki–Trotter decomposition of the time‐evolution operator but also fixes the number of collisions, i.e., $t=n\tau$, which determines the quantum circuit depth and directly influences the simulation results. 
A larger interaction time $\tau$ reduces quantum circuit depth but amplifies the truncation error in the operator expansion and Suzuki–Trotter decomposition, whereas a smaller $\tau$ suppresses this error at the cost of deeper circuits. 
It is thus essential to systematically explore various evolution operator mapping schemes and $\tau$ values.

Therefore, we introduce an error function to quantitatively compare the three schemes. The error is defined as 
\begin{equation}
  \Delta_{\text{error}}(n)=\sum_{\alpha=x,y,z}|\text{tr}[\sigma^{\alpha}_{S}\varrho_{\text{QCM}}^{n}]-\text{tr}[\sigma^{\alpha}_{S}\varrho_{\text{ME}}(t)]|,  
\end{equation}
where $\sigma^{\alpha}_{S}\, (\alpha=x,y,z)$ are the Pauli matrices. $\varrho^{n}_{\text{QCM}}$ and $\varrho^{n}_{\text{ME}}$ denote the system's density matrix 
after $n$ collisions in the quantum collision model and the solution of the Lindblad master equation at $t=n\tau$, respectively.

Figure~\ref{Fig:CMscheme} shows the error function as a function of the collision number $n$ under different values of decay rate $\gamma$ and interaction time $\tau$. 
The red solid, green dashed, and blue dotted lines represent the results obtained by Scheme I, II, and III, respectively.
In the weak-decay regime ($\gamma=1$), panels (a) and (b) demonstrate that Scheme I consistently exhibits the largest error, whereas Scheme II yields the smallest.
This result is expected, since Scheme I is based on the first‐order Suzuki–Trotter decomposition and therefore incurs larger truncation errors.
However, in the strong-decay regime (\(\gamma=5\)), panels (c) and (d) show that Scheme III achieves the highest accuracy. 
This improvement stems from the effective interaction strength $g = \sqrt{\gamma/\tau}$;
When $g$ is large, each application of the interaction operator induces a significant state change. In the physical dynamics, system–ancillas interaction and free evolution proceed simultaneously; for small $g$, the order of these steps has little impact, but as $g$ increases, interleaving becomes crucial. 
By partitioning the free evolution into fine sub-steps and interleaving each with the interaction operator, Scheme III more faithfully captures the concurrent processes and thus delivers superior performance under strong decay rate cases.

We implement the quantum circuits following Scheme III on the state-of-the-art noisy intermediate-scale quantum device to demonstrate the effectiveness of our experimental realization.
In the main text, we have already introduced our use of a hybrid classical and quantum algorithm to simulate the time evolution of the thermal state as the initial state in an open quantum system:
(1) we first employ a classical computer to decompose the given thermal state into a linear combination of pure states, i.e., $\varrho_{\text{th}}=\sum_{j}P_{j}|\psi_{j}\rangle\langle \psi_{j}|$;
(2) we then prepare these pure states $\{|\psi_{j}\rangle\}$ on a quantum device and simulate their evolution using the quantum circuits we have designed, obtaining the results $e^{\Ls t}|\psi_{j}\rangle\langle \psi_{j}|$;
(3) finally, we reconstruct the evolution of the thermal state by combining the measurement results from the quantum device, $\varrho_{\text{th}}(t)=\sum_{j}P_{j}e^{\Ls t}|\psi_{j}\rangle\langle \psi_{j}|$.

\begin{figure}[!tpb]
    \centering
    \includegraphics[width=0.95\textwidth]{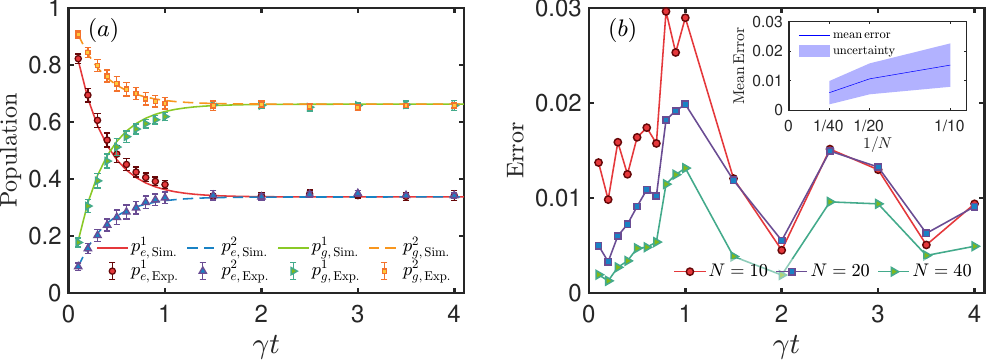}    
    \caption{\label{Fig:Exp} (a) The simulation results obtained by the quantum collision model (lines) and experimental data obtained by the quantum circuits performed on the state-of-the-art noisy intermediate-scale quantum device (markers with error bars) as a function of the time $\gamma t$. The experiment data are obtained through experimental repetitions $N=40$.  (b) The error between the simulation and experimental results as a function of the time $\gamma t$ for experimental repetitions $N=10$ (red circles), $N=20$ (blue squares), and $N=40$ (green triangles). The inset shows the mean error and its uncertainty versus the experiment repetition number $N$. For the classical simulation, the parameters are $g\tau^2=5$ and $\tau=0.02$; for the experiments, it still holds $g\tau^2=5$, but the collision time durations are $\tau_{1}=0.02$ for $\gamma t \leq 1$ and $\tau_{2}=0.1$ for $\gamma t >1$. }
\end{figure}

Figure~\ref{Fig:Exp}(a) shows the simulation results obtained by the quantum collision model (lines) and experimental data obtained by the quantum circuits performed on the quantum device (markers with error bars) as a function of the time $\gamma t$.
We measured the probability of state $|0\rangle$ and state $|1\rangle$ appearing with $\sigma_{z}$ as the basis, which corresponds to the population of the ground state $p_{g}$ and the excited state $p_{e}$, respectively.
Since we are considering a qubit system, the thermal state is decomposed into two pure states, and we denote $p_{g,\text{Exp.}}^{1}$ and $p_{g,\text{Sim.}}^{1}$ as the population of the ground state obtained by the experiment and simulation when the initial state is $|\psi_{1}\rangle$, respectively.
In addition, here we primarily aim to demonstrate the effectiveness of quantum circuits and algorithms while accounting for the hardware computing resources, we select two time durations: $\tau_{1}=0.02$ when $\gamma t \leq 1$ and $\tau_{2}=0.1$ when $\gamma t > 1$. 
This allows us to simulate a longer time evolution with fewer collisions and preserve the evolutionary details of the initial stage.
The results shown in Fig.~\ref{Fig:Exp}(a) indicate that the experimental data align well with the simulation results, demonstrating the effectiveness of our quantum-circuit implementation.
Moreover, to quantify the accuracy of our experimental results, we show the error of the reconstructed density matrix's population between the simulation and experimental results as a function of the time $\gamma t$ for different experimental repetitions $N$ in Fig.~\ref{Fig:Exp}(b). The error is defined as
$|p_{g,\text{Exp.}}(t)-p_{g,\text{Sim.}}(t)|+|p_{e,\text{Exp.}}(t)-p_{e,\text{Sim.}}(t)|$, with $p_{g(e),\text{Exp.}}(t)=\sum_{j}P_{j}p_{g(e),\text{Exp.}}^{j}(t)$ and $p_{g(e),\text{Sim.}}(t)=\sum_{j}P_{j}p_{g(e),\text{Sim.}}^{j}(t)$ being the population of the ground (excited) state obtained by the experiment and simulation, respectively.
As the number of experimental repetitions $N$ increases, the error decreases.
The inset of Fig.~\ref{Fig:Exp}(b) shows the mean error and its uncertainty (standard deviation) versus the experiment repetition number $N$, which further confirms that increasing the number of experimental repetitions can effectively reduce the error.

In summary, the results presented in Fig.~\ref{Fig:Exp} demonstrate the effectiveness of our quantum-circuit implementation and the hybrid classical-quantum algorithm in simulating the dynamics of an open quantum system with a thermal-state initial condition.

\bibliography{filterref}